# A Radio Observatory on the Lunar Surface for Solar studies (ROLSS)

R. J. MacDowall[1,6], T. J. Lazio[2,6], S. D. Bale[3,6], J. Burns[4,6], W. M. Farrell[1,6], N. Gopalswamy[1], D. L. Jones[2,6], and K.W. Weiler[5]

(1) NASA/GSFC, Greenbelt, MD, USA, (2) JPL, Pasadena, CA, USA, (3) SSL, Berkeley, CA, USA, (4) U. Colorado, Boulder, CO, USA, (5) NRL, Washington, DC, USA, (6) NASA Lunar Science Institute, NASA Ames Research Center, CA, USA

## 1. Introduction

By volume, more than 99% of the solar system has not been imaged at radio frequencies. Almost all of this space (the solar wind) can be traversed by fast electrons producing radio emissions at frequencies lower than the terrestrial ionospheric cutoff, which prevents observation from the ground. To date, radio astronomy-capable space missions consist of one or a few satellites, typically far from each other, which measure total power from the radio sources, but cannot produce images with useful angular resolution. To produce such images, we require arrays of antennas distributed over many wavelengths (hundreds of meters to kilometers) to permit aperture synthesis imaging. Such arrays could be free-flying arrays of microsatellites or antennas laid out on the lunar surface. In this white paper, we present the lunar option. If such an array were in place by 2020, it would provide context for observations during Solar Probe Plus perihelion passes. Studies of the lunar ionosphere's density and time variability are also important goals.

This white paper applies to the Solar and Heliospheric Physics study panel.

## 2. Scientific Goals

High-energy particle acceleration occurs in diverse astrophysical environments including the Sun and other stars, supernovae, black holes, and quasars. A fundamental problem is understanding the mechanisms and sites of this acceleration, in particular the roles of shock waves and magnetic reconnection. Within the inner heliosphere, solar flares and shocks driven by coronal mass ejections (CMEs) are efficient particle accelerators.

Low frequency observations provide an excellent remote diagnostic because electrons accelerated in these regions can produce intense radio bursts. The intensities of these bursts make them easy to detect, as well as providing information about the acceleration regions. The radio burst mechanisms discussed here occur at the local plasma frequency, $f_p \approx 9\, n_e^{1/2}$ kHz, or its harmonics, where $n_e$ is the electron density in cm$^{-3}$. With a model for $n_e$, $f_p$ can be converted into a height above the corona, and changing $f_p$ can be converted into radial speed. Observations by widely-separated spacecraft permit triangulation.

Solar radio bursts are one of the primary remote signatures of electron acceleration in the inner heliosphere and our focus is on two emission processes, referred to as Type II and Type III radio bursts. Type II bursts originate from suprathermal electrons (E > 100 eV) produced at shocks. These shocks generally are produced by CMEs as they expand into the heliosphere with Mach numbers greater than unity. Emission from a Type II burst drops slowly in frequency as the shock moves away from the Sun into lower density regions at speeds of 400–2000 km s$^{-1}$. Type III bursts are generated by fast (2–20 keV) electrons from magnetic reconnection, typically due to solar flares. As the fast electrons escape at a significant fraction of the speed of light into the heliosphere open along magnetic field lines, they produce emission that drops rapidly in frequency (see Figure 1).

Electron densities in the inner heliosphere yield relevant frequencies below ~10 MHz. Observations must be conducted from space because the ionosphere is opaque in this frequency range. Figure 1 illustrates the active low-frequency radio environment in space, including terrestrial radio frequency interference (RFI), as seen by the WAVES instrument on the Wind spacecraft (Bougeret et al. 1995). Solar radio observations from the moon would necessarily take place in the gaps between the RFI.

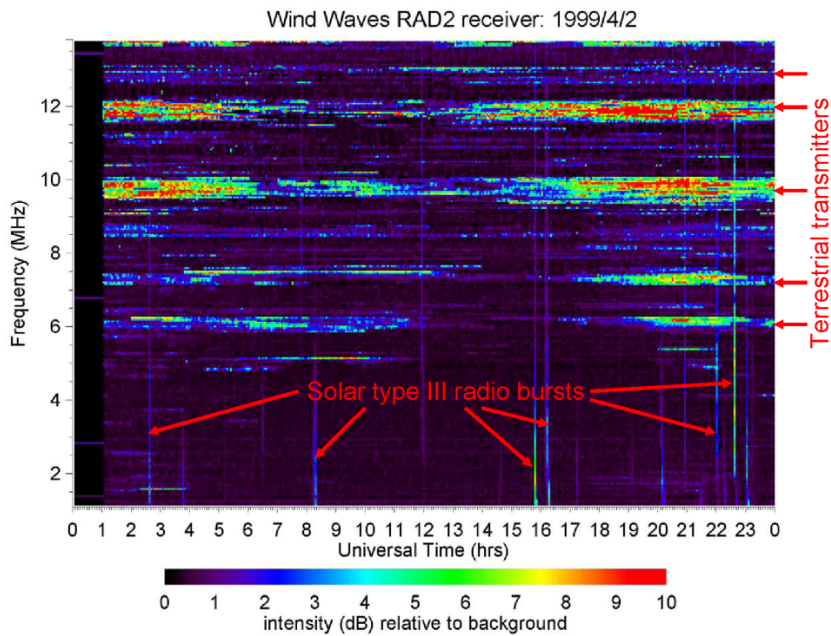

Fig. 1 - Observed low-frequency radio environment when the Wind spacecraft was near the moon. The 24-hr dynamic spectrum covers 1–14 MHz, with the intensity on a logarithmic color scale.

**Acceleration at Shocks:** Observations of CMEs near Earth suggest electron acceleration generally occurs where the shock normal is perpendicular to the magnetic field (Bale et al. 1999), similar to acceleration at planetary bow shocks and other astrophysical sites. This geometry may be unusual in the corona, where the magnetic field is largely radial, as shown schematically in Figure 2a. There, the shock at the front of a CME generally has a quasi-parallel geometry (Q-∥). Acceleration along the flanks of the CME, where the magnetic field-shock normal is quasi-perpendicular (Q-⊥) would seem to be a more likely location for the electron acceleration and Type II emission. The radio array needs ~2º resolution to localize these acceleration site(s), yielding the preferred geometry (Q-∥ vs. Q-⊥) for radio emission around CMEs.

**Electron and Ion Acceleration:** Observations at 1–14 MHz made with the Wind spacecraft showed that complex Type III-L bursts are highly correlated with CMEs and intense (proton) solar energetic particle (SEP) events observed at 1 AU (Cane et al. 2002; Lara et al. 2003, MacDowall et al. 2003). While the association between Type III-L bursts, proton SEP events, and CMEs is now secure, the electron acceleration mechanism remains poorly understood. Two competing sites for the acceleration have been suggested: at shocks in front of the CME or in reconnection regions behind the CME; see Figure 2b. For typical limb CMEs, the angular separation of the leading edge of the shock and the hypothesized reconnection region behind the CME is approximately 1.5° when the CME shock is 3–4 $R_\odot$ from the Sun.

**CME Interactions and Solar Energetic Particle (SEP) Intensity:** Unusually intense radio emission can occur when successive CMEs leave the Sun within 24 hours, as if CME interaction produces enhanced particle acceleration (Gopalswamy et al. 2001, 2002). Statistically associated with intense SEP events (Gopalswamy et al. 2004), this enhanced emission could result from more efficient acceleration due to changes in field topology, enhanced turbulence, or direct interaction of the CMEs. Lack of radio imaging makes it difficult to determine the nature of the interaction. Images with ~2 degree resolution would give Type II locations and permit identification of the causal mechanism and the relation to intense SEPs.

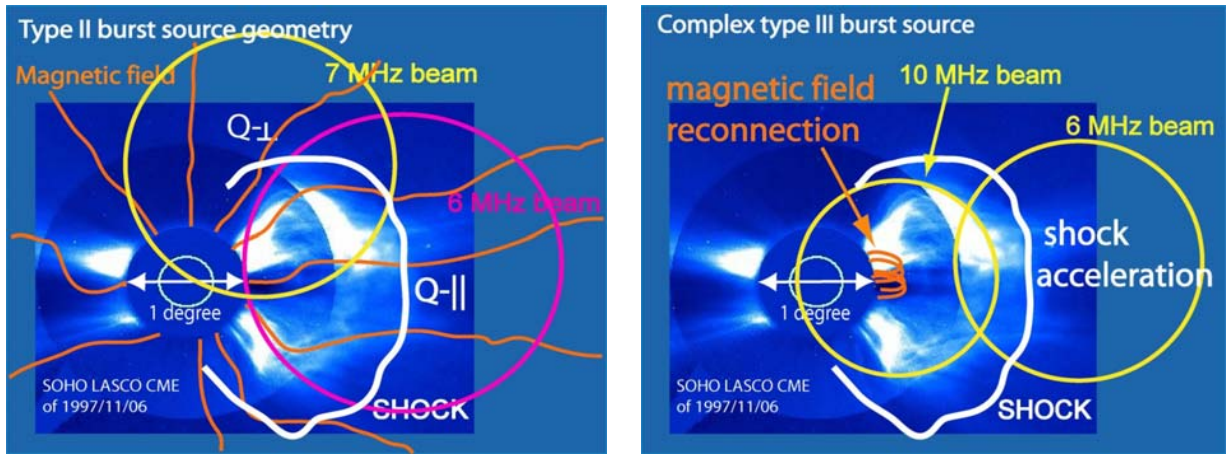

Fig. 2a – (Left) Where on the shock does electron acceleration occur, yielding type II radio emission?
Fig. 2b - (Right) Are complex type III-L bursts produced by shock acceleration or reconnection?

**Relevance to Solar Probe Plus (SPP)**: The ROLSS mission would image radio sources in the volume from ~2 to 10 solar radii, corresponding to the region sunward of SPP perihelion passes. Remote radio imaging would give locations related to solar flaring and (fast) CMEs in the environment of SPP, yielding a context for understanding solar energetic particle acceleration - one of SPP's key goals. These observations would provide a large scale perspective, complementary to those of the all-sky imager.

**Lunar ionosphere**: The lunar ionosphere is a dynamic component of the tenuous lunar atmosphere, and one that can have a significant effect on plans to use low frequency imaging arrays to study solar and interplanetary radio sources. The existing data, mainly from dual-frequency radio occultation measurements, suggest that the density of free electrons is highly variable but can exceed 2000 cm$^{-3}$ (Figure 3). The implied local plasma frequency is about 0.4 MHz.

The interpretation of these data is model dependent. Bauer (1996) concluded that the Luna data were consistent with no significant lunar ionosphere. However, ALSEP measurements during Apollo found a photoelectron layer near the surface with electron densities up to $10^4$ cm$^{-3}$ (Reasoner & O'Brien 1972), implying a plasma frequency of 0.9 MHz. Such a high plasma frequency could affect radio imaging at frequencies up to a few MHz.

Below the plasma frequency, radio waves cannot propagate. Consequently, as ROLSS tracks solar radio bursts to lower frequencies, it will also be able to search for a lunar ionosphere. If a low frequency cutoff (e.g., around a few MHz) in the radio burst spectra were observed, this would be an indication of lunar ionospheric absorption and would place a constraint on the total electron column of the lunar ionosphere.

### 3. Implementation

**Overview of the ROLSS concept:** The array consists of 3 arms arranged in a Y configuration,

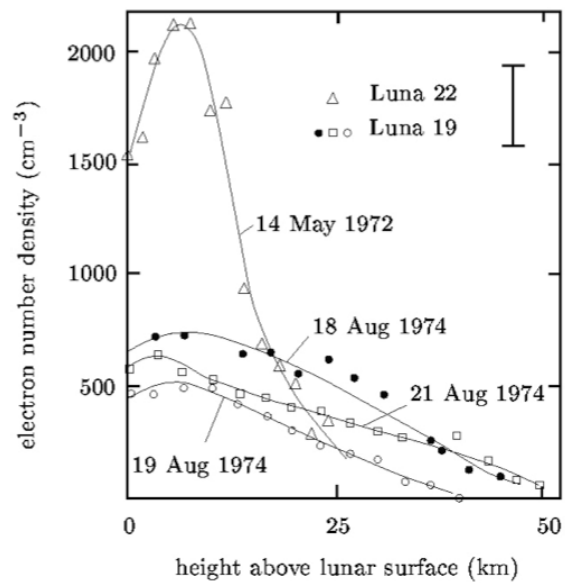

Fig. 3 - Lunar ionosphere electron densities derived from dual-frequency radio occultation measurements during the Luna 19 and Luna 22 missions (Vyshlov 1976; Vyshlov & Savich 1978).

| Table 1 – Summary of ROLSS Parameters |||
|---|---|---|
| **Parameter** | **Value** | **Comment** |
| Wavelength (Frequency) | 30–300 m (1–10 MHz) | • Matched to radio emission generated by particle acceleration in the inner heliosphere<br>• Provide context for observations during Solar Probe Plus perihelion passes<br>• Detect lunar ionosphere<br>• Operate longward of Earth's ionospheric cutoff |
| Angular Resolution | 2º (at 10 MHz) | • Localize particle acceleration sites of CME shocks and Type III solar bursts<br>• Order of magnitude improvement from present |
| Bandwidth | 100 kHz | Track time-evolution of particle acceleration |
| Lifetime | 1 yr | Obtain measurements during several solar rotations |

subject to local topographic constraints. Each arm is 500 m long, providing ~2º angular resolution at 30-m wavelength (10 MHz). The arms themselves consist of a polyimide film (PF) on which electrically-short dipole antennas are deposited, and they hold the transmission system for sending the electrical signals back to the central electronics package (CEP), located at the intersection of the arms. The CEP performs the requisite filtering and digitization of the signals, then downlinks them to Earth for final imaging and scientific analysis.

The array operates over the wavelength range 30–300 m (1–10 MHz), with a selectable, variable frequency sub-band that can be placed anywhere within the operational wavelength range. During the course of the concept study, the NASA/GSFC Instrument Design Laboratory (IDL) provided an intensive engineering study of the ROLSS concept. We refer to the conclusions of that study as output from the "IDL run."

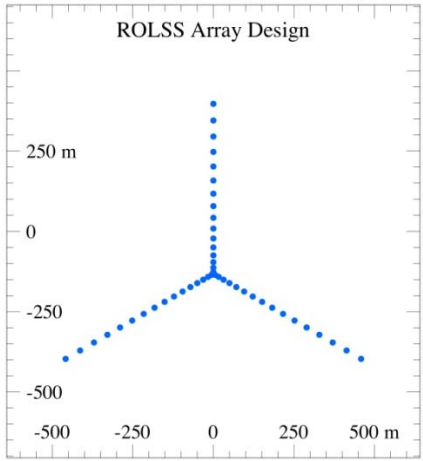 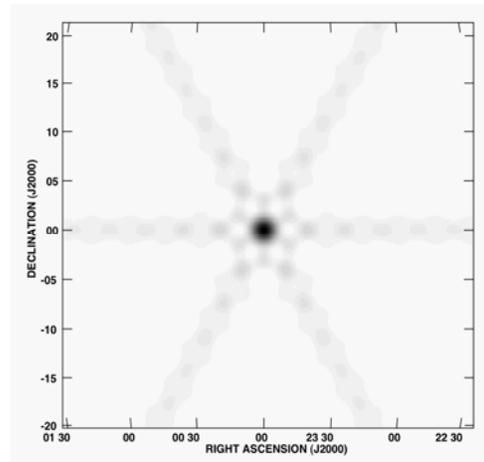

Fig. 4a - (Left) Nominal science antenna distribution along the antenna arms. Fig. 4b – (Right) The resulting point-spread function ("beam") for a snapshot image. The maximum sidelobe is at −5.9 dB and the rms sidelobe level is −15 dB, sufficient to achieve ROLSS goals.

Figure 4a shows a nominal layout of 16 science antennas along each arm. Standard radio astronomical software within the Astronomical Image Processing Software (AIPS) package of the National Radio Astronomy Observatory (NRAO) was used to simulate the instantaneous or "snapshot" point spread function (PSF) or "beam" of the ROLSS array, assuming 16 antennas on each arm and logarithmic spacing along the arm. The 6-arm "star" pattern of the beam in Figure 4b reflects the Y shape of the ROLSS array. The dynamic range in the beam - defined as the ratio between the peak and the rms level - is 15 dB. This dynamic range is consistent with a simple estimate that the rms level in an interferometric image should be of order $1/N$ if the array consists of $N$ antennas. For $N \sim 50$ ($\sim 16 \times 3$), the expected rms level is 2% (-17 dB) of the peak.

**Antenna and substrate design**: The ROLSS array consists of multiple science antennas. Each antenna is a single polarization, electrically short dipole, deposited on a PF; also deposited on the PF are the transmission leads to the CEP. The PF is flexible enough to be stored in a roll during transit and deployed directly on the lunar surface by unrolling.

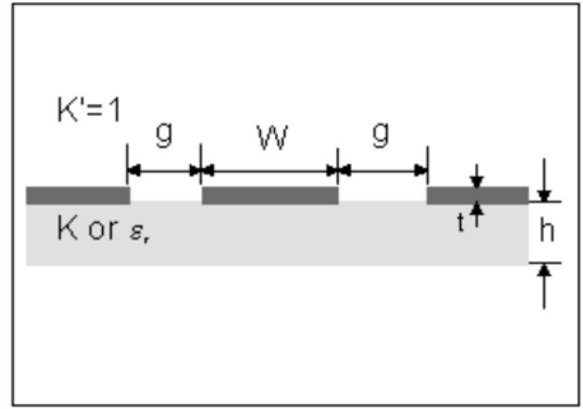

Fig. 5 - Cross section of the transmission line. The light gray indicates the substrate. The dark gray shows the deposited metal. (Figure from IDL run.)

The lunar regolith has a typical relative dielectric permittivity of 2–3 (for lunar soils), high resistivity, and a large skin depth at low radio frequencies. Consequently, it is possible for a low frequency radio antenna to operate when placed directly on the lunar surface. Modeling of the PF antennas suggest using a nominal length of 14 meters. The antenna resonant frequency will be near the high end of the ROLSS observing frequency range. At the low end of this range, the antenna will be electrically short and will have a small feed point resistance and high capacitive reactance. The radiation resistance, efficiency, and fractional bandwidth will all be very small, thereby reducing the sensitivity, but the ROLSS science goals can still be met. The baseline design has 16 antennas per arm spaced logarithmically in order to provide a range of Fourier spacings for good imaging. The nominal topology is a wide dipole, with ~1.5-m width, which helps provide a broad wavelength response.

**Signal transmission to the central hub:** Each of the dipoles is connected to the CEP at the end of the PF by transmission lines deposited on the film. Each transmission line is a coplanar waveguide structure (62 mm wide), see Figure 5. Simulations, conducted as part of the IDL run, indicate that the gap between the lines reduces cross coupling and suggest that the lines should be placed symmetrically to reduce antenna pattern distortions. These simulations suggest that ~25 dB of signal loss will occur on the longest transmission lines, which must be compensated for at the receiver. Investigation of using active elements on the PF and alternate signal transmission line designs is ongoing.

The width of the transmission lines is determined in part by the requirement of mitigating micrometeorite damage. During the IDL run, estimates of the micrometeorite damage were determined by scaling from experience with the Hubble Space Telescope (HST) Wide Field/Planetary Camera (WF/PC) radiator. The estimated bombardment rate at the Moon's surface is 4.9 $m^{-2}$ $yr^{-1}$ for impacts producing craters up to 1 mm wide. Larger craters are sufficiently rare that none were seen in a 3.6-yr exposure. The longest transmission lines are likely to receive a few impacts per year (~ 3), but none are likely to be so large that the transmission line will be completely severed.

**Preamplifiers, receiver, data processing, and data storage:** The ROLSS CEP (shown in Figure 6) houses all of the science antenna receivers including amplification, data acquisition hardware, power supply electronics, data and telemetry communications electronics, and the thermal management system for these components. When deployed, the CEP stands ~1 m above the lunar surface in order that the thermal management system is not subject to dust contamination.

**Data downlink:** ROLSS uplink telemetry and commanding would be handled by an S-band uplink. A nominal configuration has a 9-m diameter or larger antenna at a site such as GSFC Wallops. The ROLSS

CEP would contain an S-band transceiver, which could also be used for data downlink in emergency situations.

Data downlink would be handled by a Ka-band system. The ROLSS CEP would host a 0.6-m diameter parabolic gimbaled antenna. The conclusion of the IDL run was that a Ka-band downlink via such an antenna provided the current optimum based on downlink data requirements, weather (rain) conditions, and likely ground-station antenna availability.

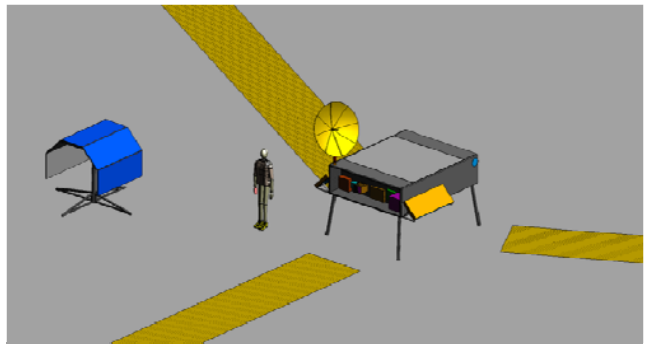

Fig. 6 – ROLSS Central Electronics Package (CEP) layout showing relative sizes of electronics box with thermal louvers, high-gain antenna, and solar arrays. (Figure from IDL run.)

**Siting and deployment:** The prime science mission for ROLSS is solar observations, which favors an equatorial site. A polar site would require that one or more arms of ROLSS be longer than the nominal 500 m length in order to compensate for the foreshortening of the array and maintain the same angular resolution.

The site itself should be of relatively low relief topography (e.g., the surface of a *mare* vs. the highlands). We emphasize that the entire site does not have to be flat as the fraction of the area occupied by the antenna arms is relatively small—for a circular area with a radius of 500 m, the antenna arms occupy only 0.3% of the total area. Further, there is no requirement on the absolute orientation of the arms. Rotating the arms about the center of the array merely rotates the beam pattern on the sky. The *relative* orientation of the arms is required to be 120º ± 6º, with the error determined by requiring that the nominal antenna position not vary by more than 1 wavelength at the extreme end of an arm. This level of alignment was demonstrated during the Apollo missions. Only modest requirements are set on the smoothness of the antenna arm locations. The shortest operational wavelength for ROLSS would be ~30 m, and an individual science antenna would be 15 m in length. Requiring that the antenna shapes are not distorted by more than $\lambda/10$ implies that deployment requires linear extents of order 500 m with elevation variations not larger than ~3 meters along the intended antenna arm positions. We assume that on the timescale of likely deployment for ROLSS that any site will have better than meter-scale resolution images available (post-Lunar Reconnaissance Orbiter).

The ROLSS deployment itself could be done completely robotically, by astronauts on a crewed rover, or with a mix of these two modes. A completely robotic deployment would allow ROLSS to be deployed prior to any human sortie missions. Scaling from existing rovers, a ROLSS rover could easily be designed to carry and deploy tens of kilograms of antenna mass.

**Other ROLSS concept elements:** The ROLSS mission concept calls for lunar day operations only; there are no nighttime operations. The baseline ROLSS design assumes a solar panel assembly for power generation. Multiple load profiles were generated during the IDL run, with the final load profile suggesting that an average 125 W load (peak 167 W load) would need to be supported by solar panels.

Thermal management is a key requirement for lunar-based instrumentation. The antenna arms are entirely passive and require no thermal management. The CEP is equipped with a set of radiators in order to maintain the internal electronics at a temperature below their assumed maximum operating temperature (80º C). These radiators are equipped with thermal louvers, outfitted with actuators that sense radiator temperature and rotate thin aluminum blades to open or closed positions. When open, the louvers allow the radiators to dissipate heat. During lunar night, batteries provide survival heating.

**Other radio imaging options:** In this white paper, we have focused on the polyimide film antenna system, but other viable approaches for antennas on the lunar surface are also being studied. Also, a constellation of microsats carrying dipole antennas is an alternative for low-frequency radio imaging (MacDowall et al. 2006).

## 4. Cost Estimate

The IDL run did not involve cost estimation, but did provide masses for the ROLSS components, assuming implementation using present day technology. In the table below, we list the components, estimated masses assuming additional technology development, and ROM cost estimates for a mission similar to the IDL run, with the modification of deployment by robotic rover. This well-documented, but conservative, concept design would be easy to price. Its major weakness is high mass. Alternatively, additional technology development could result in mass savings using advanced techniques like a phased array for the downlink, ultra-low temperature/ultra-low power electronic components, and advanced battery, thermal, and power components. The mass and cost of the rover are likely to be inversely related, with reliability also entering into the trade space.

| Component | Mass Estimate (kg) | ROM Cost Estimate (M$) | Details of ROM estimate |
|---|---|---|---|
| Science antenna array, including development | 90 | 10 | Primarily development cost |
| Central Electronics Package | 140 | 12 | Development cost |
| Lithium Ion Batteries | 120 | 5 | Hardware cost |
| CEP Thermal system | 25 | 3 | Development & hardware cost |
| RF/Comm System | 25 | 5 | Hardware cost |
| Solar Panel Assembly | 25 | 2 | Hardware cost |
| Deployment rover system | 50 | 20 | Hardware & software development + testing |
| Operations | n/a | 12 | Labor cost + facilities |
| Systems Engineering | n/a | 4 | Labor cost |
| Project Management | n/a | 16 | Based on costs for other missions |
| Total cost excluding LV and lander | 475 | 89 | |

## 5. Conclusion

Investigation of particle acceleration and propagation using radio bursts as diagnostic tools has been performed to date without imaging capability below ~20 MHz. A radio observatory such as ROLSS, located on the lunar surface, would provide the first images of radio bursts in the outer solar corona and interplanetary space, yielding new information on solar flaring and shock acceleration of particles. ROLSS observations would be of particular interest to establish the context for perihelion data from the SPP mission. The complement of ROLSS images, white light data from the SPP all-sky imager, and ground-based solar radio observations would constrain the large-scale space weather environment in the vicinity of SPP and sunward of the spacecraft.